\documentstyle[12pt]{article}
\begin{document}
\baselineskip=26pt
\begin{flushright}
OCHA-PP-75
\end{flushright}
\vskip 0.5 cm
\begin{center}
{\Large \bf Induced Light-Quark Yukawa Couplings \\
as a probe of Low-Energy Dynamics in QCD}
\vskip 1cm

{\bf  M. R. Ahmady}\\
Department of Physics, Ochanomizu University, \\
 1-1, Otsuka 2, Bunkyo-ku, Tokyo 112, Japan.\\
\vskip 1.0cm
{\bf V. Elias, A. H. Fariborz, R. R. Mendel}\footnote{Deceased}\\
Department of Applied Mathematics, The University of Western Ontario, \\
 London, Ontario, Canada N6A 5B7. \\ 
\vskip 1cm
April 1, 1996
\vskip 1cm       
{\bf Abstract}
\vskip 0.5cm       
\end{center}         

We examine the heavy-quark-induced Yukawa interaction between light 
quarks and a light Higgs field, which is facilitated in the chiral 
limit by the dynamical mass of light quarks anticipated from the 
chiral-noninvariance of the QCD vacuum.  A low-energy estimate of the 
strong coupling near unity can be obtained from a comparison of the 
explicit perturbative calculation of the induced Yukawa interaction 
at zero momentum to a Higgs-low-energy theorem prediction for the same 
interaction.

\newpage

      The effective coupling of zero-momentum Higgs particles to
nucleons is well-known \cite{HNCoupling,DawHab} to be approximately
of magnitude $70 MeV N_h/\langle \phi
\rangle $, where $N_h$ is the number of heavy quark flavours, and where
$\langle \phi \rangle $ is the vacuum expectation value of the Higgs field. 
Although this result is physical only if the mass of the Higgs is
light compared to $F_{ew}$, the electroweak scale, we can
nevertheless use it at the quark level to probe the strong coupling
at low energies.

In this paper we consider the coupling of a zero momentum Higgs
particle to light quarks to get insight into the magnitude of the 
low-energy QCD coupling.  In other words, we augment QCD with a light
({\it gedanken}) Higgs boson in order to explore nonperturbative QCD dynamics.  
Specifically, we compare Higgs-low-energy
theorem predictions appropriate for light quarks to the  explicit
perturbative calculation of the Yukawa interaction induced via 
heavy-quarks.  This comparison provides an estimate of the QCD coupling
characterizing the latter calculation.  In both cases, we work in the chiral
limit in which the light quark is assumed to have no tree-level Yukawa
interaction, but is assumed to obtain mass dynamically via the chiral
noninvariance of the QCD vacuum \cite{Higashijima,CNQCDV,ChiralSB}.  For 
the low-energy theorem
calculation of the light-quark Yukawa interaction, this assumption implies
an effective quark mass proportional to $\Lambda _{QCD}$, leading to a 
calculation identical in form to the low-energy-theorem prediction 
\cite{DawHab} for the Higgs-nucleon interaction.  For the explicit 
perturbative calculation, we consider a mass function suggested by 
Holdom \cite{Holdom} that falls off
with momentum with asymptotic $\Lambda ^3/p^2$ behaviour.  The comparison of the
quark-level low-energy theorem to the explicit quark-level calculation
suggests a low energy value for $\alpha _s$ near unity, as anticipated from 
chiral
symmetry breaking \cite{Higashijima,ChiralSB} and other 
low-energy considerations
\cite{LEStudy}. 

     To begin the perturbative quark-level calculation, we first consider
the coupling of the zero 4-momentum Higgs field to two gluons that is
induced by a heavy quark triangle (Figure 1), with p corresponding to the
external gluon momentum \cite{triangle}.  The triangle is given by
the following tensor:

\begin{equation}
I_{\mu \nu }^{ab}=2 c \delta ^{ab} Tr \int d^n k 
{
{ (k \! \! \! / +M)\gamma _\mu (k \! \! \! / - p \! \! \! / + M) ^2 
\gamma _\nu }
\over
{(M^2-k^2)\left[ M^2 -(k-p)^2 \right] ^2}
},
\label{eq:Imunug}
\end{equation} 
with
\begin{equation}
c={{-i\alpha _s M}\over {8 \pi^3  \langle \phi \rangle }},                    
\label{eq:c}
\end{equation}
In (\ref{eq:Imunug}) and (\ref{eq:c}), $M$ is the heavy 
quark mass, $\alpha_s $ is
$g_s^2/(4\pi)$, and $\langle \phi \rangle $ is the
vacuum expectation value of the Higgs field.  Thus, the parameter $c$
contains the dimensionless Yukawa coupling between heavy quarks and
the Higgs field.  The factor of 2 preceding the integral in
(\ref{eq:Imunug}) accounts for
the two possible directions for the fermion arrows in Fig. 1.   
      The denominator in (\ref{eq:Imunug}) can be combined via the usual Feynman
parameterization, so as to obtain $(k' \equiv  k - px, p^2 \leq  0)$
\[
I_{\mu \nu}^{ab}(p)=
16 M c \delta ^{ab} \int d^n k' \int _0^1  dx 
{ 
{4k'_\mu k'_\nu +4 p_\mu p_\nu (x^2-x)+(p^2+M^2-p^2x^2-k'^2)g_{\mu \nu}}
\over 
{\left[ M^2+p^2(x^2-x)-k'^2 \right] ^3 }
},
\]
\begin{equation}
=
{
{2 \alpha}
\over 
{\pi \langle \phi \rangle }
}
\delta ^{ab} (p_\mu p_\nu -p^2 g_{\mu \nu }) 
\left[ 
{M^2 \over p^2} +       {
                            {2M^4} 
                            \over 
                            {p^4 \sqrt { 1- 
                                             {
                                                 4 M^2 
                                                 \over
                                                 p^2 
                                             }
                                       }
                            }
                        }
                          ln | {{\tau _+} \over {\tau _-}}|
\right],
\label{eq:Imunusolution} 
\end{equation}
with
\begin{equation}
\tau _{ \pm } 
\equiv 
1 \pm 
\sqrt { 1- 
            {
                4 M^2
                \over
                p^2 
            }
      }.
\end{equation}
The transversality of this expression ensures gauge-parameter
independence of light-quark Yukawa couplings induced via Figure 2.
Two limits of interest are the limits of infinite mass and infinite
momentum:
\begin{equation}
\lim_{M\rightarrow \infty } I_{\mu \nu }^{ab}(p)=
{
   {-\alpha}  
   \over
   { 3 \pi \langle \phi \rangle }
}
\delta ^{ab} (p_\mu p_\nu -p ^2g_{\mu \nu }),
\label{eq:ImunulimitM}
\end{equation}
\begin{equation}
\lim_{p^2 \rightarrow -\infty }
\left[ 
{M^2 \over p^2} +       {
                            {2M^4} 
                            \over 
                            {p^4 \sqrt { 1- 
                                             {
                                                 4 M^2 
                                                 \over
                                                 p^2 
                                             }
                                       }
                            }
                        }
                          ln | {{\tau _+} \over {\tau _-}}|
\right] = 0.
\label{eq:Imunulimitp}
\end{equation}

      Equation (\ref{eq:ImunulimitM}) illustrates the well-known
result that for very heavy quarks, the quark triangle
is independent of the heavy quark mass \cite{HNCoupling,DawHab}, 
suggesting that heavy flavours
c, b, and t will generate equivalent contributions to the Yukawa interaction
induced via Figure 2.  This property ensures that the contribution of
Figure 2 to the light-quark Yukawa interaction will be proportional to the
number of heavy flavours, as anticipated from the Higgs-nucleon
interaction discussed at the begining of this paper.  Substitution of 
(\ref{eq:Imunulimitp}) into (\ref{eq:Imunusolution}) demonstrates 
that the quark triangle will serve 
as a cut-off in integrals over the gluon momentum of Figure 2, 
ensuring a finite renormalization-independent result. 
In other words, the induced Yukawa interaction is an infrared effect.

The Yukawa coupling  to light quarks induced via Figure 2 may be 
expressed as follows:
\[
\Sigma _{ind}(k) = \int {d^4p \over {i (2\pi )^4 }}
g_s T_a \gamma ^\mu
{
  1
  \over 
  {-(k \! \! \! / - p \! \! \! / )+ \Sigma _{QCD} [(k-p)^2]}
}
g_s T_b \gamma ^{\nu } 
{1\over p^2} I_{\mu \nu }^{ab}(p) {1\over p^2 },
\]
\begin{equation}
\equiv C(k^2)k \! \! \! / + B(k^2).
\label{eq:Sigmaind}
\end{equation}
To evaluate this expression, we find it convenient to express the 
heavy quark triangle 
(\ref{eq:Imunusolution}) in the
following form:
\begin{equation}
I_{\mu \nu }^{ab}(p)= 
{
   {4 \alpha _s M^2 }
   \over
   {\pi \langle \phi \rangle }
}
\delta ^{ab} (p_\mu p_\nu -p^2 g_{\mu \nu }) 
\int _0^1 dy 
{
   y
   \over
   { p^2 - 
          {
            M^2
            \over
            {y(1-y)}
          }           
    }
}. 
\label{eq:Imunusimple}
\end{equation}
Let us first suppose that the light quark's mass term is constant.  In this
case we must replace the factor $\Sigma_{QCD}[(k-p)^2]$ 
in the denominator of the
integrand in (\ref{eq:Sigmaind}) with a hard mass 
denoted by $m_{dyn}$.  We then find that the
induced Yukawa interaction is given by 
\begin{equation}
B(k^2) = 
{
   {4 \alpha ^2 m_{dyn} }
\over 
{\pi ^2 \langle \phi \rangle }
} 
\int _0^1 dy y^2(1-y) \int _0^1 dx_1 \int_0^{1-x_1} dx_2 
{
1
\over
{x_1+y(1-y)x_2\left[ {m_{dyn}^2\over M^2} +{k^2 (1-x_2) \over M^2} \right] }
},                                                        
\end{equation}
and for $m_{dyn} \ll M$ 
\begin{equation}
B(0) \rightarrow 
{ 
   {2 \alpha _s^2 }
   \over
   {3\pi ^2 \langle \phi \rangle }
}
m_{dyn}
\left[ 
  ln (
      {
         M
         \over
         m_{dyn}
      }
     )
+{5\over 6}
\right]. 
\label{eq:B0limit}
\end{equation}
This induced Yukawa interaction is seen to be characterized by a mass
that is suppressed by a factor of $2\alpha _s^2 [ln(M/m_{dyn})+
5/6]/(3 \pi ^2)$ relative to the hard
(dynamical) light-quark mass.  However, this estimate tells us
very little--the presence of the large 
logarithm in (\ref{eq:B0limit}) necessarily
indicates sensitivity of the Feynman integral to the ultraviolet region, in
which case one cannot assume that the factor $\alpha_s^2$ appearing
in (\ref{eq:B0limit}) characterizes exclusively low-energy dynamics. 

      We now consider the more realistic case in which
$\Sigma_{QCD}[(k-p)^2] $ is interpreted as a dynamical mass function 
generated by the chiral noninvariance of the QCD vaccuum.  
This self-energy is expected to fall  off like $1/(k-p)^2$ for large 
$|k-p|$.  Since this dynamical mass decreases sharply 
with increasing momentum, the Yukawa interaction induced via Figure
2 is more sensitive to the infrared portion of the integration region, 
corresponding to the low-energy region for $\alpha_s$. We utilize 
the phenomenological 
self-energy proposed by Holdom \cite{Holdom}
\begin{equation}
\Sigma_{QCD}(p^2)=
{
   (A+1)\Lambda^3 
   \over 
   A\Lambda^2-p^2
},
\end{equation}
in conjunction with (\ref{eq:Imunusimple}), in order to obtain the 
following expressions for
$C(k^2)$ and $B(k^2)$ in (7):
\begin{equation}
C(k^2) k \! \! \! / = 
{
{4 \alpha_s ^2 M^2}
\over 
{3i\pi^4 \langle \phi \rangle }
}
\int _0^1 ydy \int d^4p 
{
   {\gamma_\mu (p\! \! \! / -k\! \! \! /)
   \left[A \Lambda^2 -(p-k)^2\right]^2\gamma_\nu (p^\mu p^\nu-p^2g^{\mu
   \nu } )}
   \over 
   {
    \left[ 
          (p-k)^2 \left[ A\Lambda^2 -(p-k)^2 \right]^2
          -(A+1)^2\Lambda^6
    \right] 
            p^4 
                 \left( p^2-  {
                                M^2 
                                \over 
                                y(1-y)
                               }
                 \right) 
   } 
},                                                               
\end{equation}
\begin{equation}
B(k^2) = 
{
{4 \alpha_s^2 M^2}
\over 
{3i\pi^4\langle \phi \rangle }
}
\int_0^1 y dy \int d^4p 
{ 
{\gamma_\mu  (A+1)\Lambda^3  \left[ A\Lambda^2-(p-k)^2 \right]
\gamma_\nu (p^\mu p^\nu -p^2g^{\mu \nu})  
} 
\over
{\left[ (p-k)^2 \left[ A\Lambda^2 -(p-k)^2 \right] ^2 -(A+1)^2\Lambda
^6 \right] p^4 \left( p^2 - {M^2 \over {y(1-y)}} \right)}              
}.
\end{equation}
We can easily evaluate these integrals numerically when $k^2 = 0$, 
the quark's Lagrangian-mass shell in the chiral limit.  Upon performing the Wick
rotation $d^4p \rightarrow id^4p_E = i\pi^2 x dx$, we find that
\begin{equation}
C(0) = 
{
{2 \alpha ^2  M^2 (A+1)^2 \Lambda ^6 } 
\over
{\pi^2 \langle \phi \rangle}
}
\int_0^1 y dy \int _0^{\infty } dx
{
{3 x^2 + 4 A \Lambda ^2 x + A^2 \Lambda ^4 }
\over 
{\left[ x ( A \Lambda ^2 +x )^2 + ( A + 1 )^2 \Lambda ^6 \right] ^2
\left( x + {M^2 \over {y(1-y)}} \right) }
},
\label{eq:C0}
\end{equation}                                                         
\begin{equation}
B(0) = 
{
{4 \alpha ^2  M^2 (A+1) \Lambda ^3 } 
\over
{\pi^2 \langle \phi \rangle}
}
\int_0^1 y dy \int _0^{\infty } dx
{
{A \Lambda ^2 + x }
\over 
{\left[ x ( A \Lambda ^2 +x )^2 + ( A + 1 )^2 \Lambda ^6 \right] 
\left( x + {M^2 \over {y(1-y)}} \right) }
}.
\label{eq:B0}
\end{equation}                                                         
Choosing $A = 2$ and $\Lambda = 317 MeV$ \cite{Holdom}, we find that 
\begin{equation}
B(0) = 62 \alpha_s ^2 MeV / {\langle \phi \rangle}.
\label{eq:B0A2L317}
\end{equation} 
In the chiral limit, there is no Lagrangian mass term for the light quark,
and the $C(k^2)k \! \! \! /$ is expected (via Lagrangian 
field equations) to annihilate
an external light quark spinor $u(k)$ when $k^2 = 0$.  However, the
contribution of $C(0)$ to the induced Yukawa interaction is seen to be
small even if we go so far as to identify $k \! \! \! /$ 
with an order-$300 MeV$
dynamical mass. Specifically, we find $C(0) = 0.051 \alpha_s^2/
\langle \phi \rangle $ when $A = 2$ and $\Lambda = 317 MeV$, in which case
$C(0)m_{dyn} [\approx 15 \alpha_s^2 MeV/\langle \phi \rangle] $ is still
substantially smaller than $B(0)$. \footnote{See N. C. A. Hill and V.
Elias \cite{HillElias} for a discussion as to the validity of
renormalization on mass shells of dynamical (non-Lagrangian)
origin.}

      Thus, we find from (\ref{eq:B0A2L317}) that the effective mass 
characterizing the induced Yukawa interaction between a zero-momentum light 
quark and a zero-momentum Higgs to be of order $60 \alpha_s^2 MeV$.  
Inclusion of a dynamical mass shell [i.e., adding $C(0) m_{dyn}$ to
$B(0)$ ] would increase this estimate to about $77 \alpha_s^2 MeV$.
We reiterate that the light quarks are assumed to possess only dynamical mass
contributions--the Lagrangian Yukawa coupling to light quarks is
assumed to be zero, consistent with the chiral limit.  

      We now compare (\ref{eq:B0A2L317}) to the Yukawa coupling anticipated
via Higgs Boson low-energy theorems.  The relevant low-energy theorem
between matrix elements $A \rightarrow B $ and $A \rightarrow B + Higgs
$ is given by \cite{DawHab}
\begin{equation}
\lim_{p_{Higgs} \rightarrow 0} M(A \rightarrow B + Higgs )
=
{ 
   N_h \alpha_s ^2
   \over
   3\pi \langle \phi \rangle 
}
{
   \partial 
   \over
   \partial \alpha_s
}
M ( A \rightarrow B ), 
\label{eq:LET}
\end{equation}
where $N_h$ is the number of heavy quarks.  If A and B are both identified
with constituent u or d quarks, then the physical
matrix element $M(A\rightarrow  B) $ is just the dynamical quark mass
$m_{dyn}$, which is clearly proportional to $\Lambda_{QCD}$. To one loop order, 
\begin{equation}
\alpha_s = 
{
   {(\pi d)}
   \over 
   {ln ({p^2\over \Lambda_{QCD}^2}) }
}, 
\end{equation}
with $d = 4/9$ for three light quark flavours. Consequently, one finds that
if $m_{dyn} \equiv \kappa \Lambda_{QCD}$, then
\begin{equation}
{
   \partial m_{dyn}
   \over
   \partial \alpha_s 
}
= \kappa 
{
   \partial \Lambda_{QCD}
   \over
   \partial \alpha_s 
}
=
{
   {\kappa (\pi d) \Lambda_{QCD}}
   \over
   {2 \alpha_s^2}
}
=
{
   (\pi d) m_{dyn}
   \over
   2 \alpha_s^2
}.
\end{equation}
Substitution of this result into the low-energy theorem for light-quark
$(q)$ higgs couplings yields the result
\begin{equation}
\lim_{p_{Higgs} \rightarrow 0} M(q \rightarrow q + Higgs )
\equiv
g_{Hqq}|_{induced}
=
{
   2 N_h m_{dyn}
   \over
   27 \langle \phi \rangle 
},
\label{eq:LETlq}
\end{equation}
corresponding, for $N_h = 3 $ (c,b,t) to a Yukawa coupling characterized by
a mass of about $70 MeV$.\footnote {This is precisely the same
mechanism used by Dawson and Haber \cite{DawHab} to predict the
Higgs-nucleon-nucleon coupling, as the chiral limit of the nucleon
mass is also assumed to be proportional to $\Lambda_{QCD}$. 
Replacing the dynamical quark mass with the nucleon mass in (19) leads
to Dawson and Haber's prediction for the Higgs-nucleon-nucleon
coupling of order $200 MeV/ \langle \phi \rangle $.} 

      We see that comparison of (\ref{eq:LETlq}) and
(\ref{eq:B0A2L317}) suggests a low-energy
value for $\alpha_s$ near unity, consistent with criticality arguments for 
chiral symmetry breaking \cite{Higashijima,ChiralSB} as well as with 
the freeze-out of the strong coupling
near unity discussed by Mattingly and Stevenson \cite{LEStudy}. 
There are, of course, concerns that such a large value of $\alpha_s$
will destroy the validity of the perturbative expansion.  There is
some reason to believe, however, that the relevant expansion
parameter near criticality of $\alpha_s$ is $\alpha_s N_c/ 4 \pi $
\cite{Near_Criticality}, perhaps providing some additional supression
of higher-order contributions.  Note also that a comparison 
of (\ref{eq:LETlq}) to the hard dynamical-mass result
(\ref{eq:B0limit}) for each heavy quark flavour would suggest a smaller 
value for $\alpha_s$ ($\alpha_s \approx 1/2$), but this value cannot be
exclusively identified with the low energy region of hadron physics, as has
already been discussed. 

Moreover, we see from (\ref{eq:LETlq}) that the mass
associated with the induced Yukawa coupling to light quarks via heavy
quark loops is considerably larger in magnitude than the current
(Lagrangian) mass usually assumed for light quarks.\footnote{Enhancement 
mechanisms for the current quark
mass characterizing some phenomenological applications are discussed
in ref. \cite{Elias}.}   One can also apply 
the low energy theorem (\ref{eq:LET}) to examine whether or not Yukawa 
couplings to heavy quarks $(Q)$ are similarly characterized by enhanced 
masses (masses substantially larger than those in the Lagrangian):
\begin{equation}
\lim_{p_{Higgs} \rightarrow 0} M(Q \rightarrow Q + Higgs )
\equiv
\bigtriangleup g_{HQQ}
=
{
   N_h \alpha_s^2 
   \over
   3 \pi \langle \phi \rangle 
}
{
   \partial m_Q 
   \over
   \partial \alpha_s
}.
\label{eq:LETHQ}
\end{equation}
The heavy quark mass $m_Q$ is not proportional to $\Lambda_{QCD}$, but 
depends on $\alpha_s$ via the relationship
\begin{equation}
m_Q(p^2)
=
{
   {\hat m }
   \over
   \left[
          ln \left(
                    {
                       p^2
                       \over
                       \Lambda _{QCD}^2
                     }
             \right)
   \right]^{d'}  
}
=
{\hat m} 
\left( {\alpha_s\over {\pi d}} \right)^{d'},
\label{eq:mQp2}
\end{equation}
where ${\hat m}$  is a renormalization-group invariant quark mass, and
where $d' \equiv 12/(33-2n_f)$. Substituting (\ref{eq:mQp2}) into
(\ref{eq:LETHQ}), we find that
\begin{equation}
\bigtriangleup g_{HQQ} 
=
{
   {N_h \alpha_s d' m_Q}
   \over 
   {3 \pi \langle \phi \rangle }
},
\end{equation}
corresponding to an enhancement of at most $14 \% $ over the Lagrangian Yukawa
coupling $m_Q/\langle \phi \rangle $ if $ \alpha_s \approx 1 $. 

In conclusion, we find that a comparison of the low-energy-theorem Yukawa
interaction, characterized by a mass of $ 70 MeV$, to an explicit perturbative
calculation of the same interaction, characterized approximately by a
mass between $ 60 \alpha_s^2 - 80 \alpha_s^2 \hskip .2cm MeV$, 
suggests that the strong
coupling at low energies is surprisingly close to unity.

We are grateful for extensive discussions with V. A. Miransky, and to
the Natural Science and Engineering Research Council of Canada for
financial support.  Work of M. A. is supported by the Japanese Society 
for the Promotion of Science.  We also note with sorrow that R. R. 
Mendel, who initiated this research, was actively pursuing its resolution 
at the time of his tragic death last August.

\newpage
{\bf Figure Captions:}

{\bf Fig.1} Heavy quark triangle.

{\bf Fig.2} Coupling of Higgs to light-quark via the heavy quark
triangle.


\newpage

\begin{thebibliography}{99}

\bibitem{HNCoupling} 
M. A. Shifman, A. I. Vainshtein, and V. I. Zakharov, Phys. Lett. {\bf
B78} (1978), 443.\\
A. I. Vainshtein, V. I. Zakharov, and M. A. Shifman, Sov. Phys. Ups.
{\bf 23} (1980), 429.\\
L. B. Okun, {\it Leptons and Quarks} (North Holand, Amsterdam,1984), pp.
230-231.

\bibitem{DawHab}S. Dawson and H. Haber, Int. J. of Mod. Phys. {\bf
A7} (1992), 107.

\bibitem{Higashijima}K. Higashijima, Phys. Rev. {\bf D29} (1984), 1228.

\bibitem{CNQCDV}K. Lane, Phys. Rev. {\bf D10} (1974), 2605.\\
H. D. Politzer, Nucl. Phys. {\bf B117} (1976), 397.\\
H. Pagels, Phys. Rev. {\bf D19} (1979), 3080.\\
H. Pagels and S. Stokar, Phys. Rev. {\bf D20} (1979), 2948.\\
V. Elias and M. D. Scadron, Phys. Rev. {\bf D30} (1984), 647.

\bibitem{ChiralSB}P. Fomin, V. Gusynin, V. A. Miransky and Yu. Sitenko, 
Riv. Nuovo Cim {\bf 6} (1983), 1.\\
J. E. Mandula and J. Weyers, Nucl. Phys. {\bf B237} (1984), 59.

\bibitem{Holdom}B. Holdom, Phys. Rev. {\bf D45} (1992), 2534.

\bibitem{LEStudy}A. C. Mattingly and P. M. Stevenson,
Phys. Rev. Lett. {\bf 69} (1992), 1320.

\bibitem{triangle}F. Wilczek, Phys. Rev. Lett. {\bf 39} (1977),
1304.\\
J. Ellis, M. K. Gaillard, D. V. Nanopoulos and C. T. Sachradja, Phys.
Lett. {\bf B83} (1979), 339.\\
T. Rizzo, Phys. Rev. {\bf D22} (1980), 178.

\bibitem{HillElias}N. C. A. Hill and V. Elias, Int. J. of Mod. Phys. {\bf A9}
(1994), 181.

\bibitem{Near_Criticality}T. Appelquist, J. Terning, and L. C. R.
Wijewardhana, Yale University Preprint YCTP-P2-96 [hep-ph/9602385].

\bibitem{Elias}V. Elias, Can. J. of Phys. {\bf 71} (1993), 347.


\end{thebibliography}
\end{document}